\documentclass[11pt]{article}
\usepackage{graphicx}
\usepackage{epsfig}
\usepackage{amssymb,amsmath}
\usepackage{graphicx}  
\newcommand{\be}{\begin{equation}}
\newcommand{\ee}{\end{equation}}
\newcommand{\ba}{\begin{eqnarray}}
\newcommand{\ea}{\end{eqnarray}}

\fontencoding{T1}
\fontfamily{garamond}
\fontseries{m}
\fontshape{it}
\fontsize{13}{15}
\selectfont

\begin{document}
\title{Positive Parity Scalar Mesons in the 1-2 GeV Mass Range}
\author{L.~Maiani$^a$, F.~Piccinini$^b$ A.D. Polosa$^c$, V.~Riquer$^c$\\
\small{$^a$Dip. Fisica, Universit\`a  di Roma ``La Sapienza'' and INFN, Roma, Italy.}\\
\small{$^b$I.N.F.N. Sezione di Pavia and Dipartimento di Fisica Nucleare 
e Teorica}\\ 
\small{ via A.~Bassi, 6, I-27100, Pavia, Italy.}\\
\small{$^c$INFN, Sezione di Roma 1, Roma, Italy.}}

\maketitle
\begin{abstract}
Based on the observation that $K_0(1430)$ is lighter than its $SU_3$ counterpart, $a_0(1450)$, we examine the possibility that these particles, together with $f_0(1370)$, $f_0(1500)$ and $f_0(1710)$, fill a tetraquark recurrence of the sub-GeV $0^{++}$ nonet mixed with a glueball state. We find the picture to be consistent with the known data about the three $f_0$ resonances, more than the $q\bar q$ hypothesis. Conventional spin-orbit coupling suggests the $q\bar q$, P-wave, nonet to lie around 1200 MeV. We review possible experimental indications of a scalar isovector resonance at 1.29 GeV, first observed by OBELIX in $p\bar p$ annihilation.
ROME1-1429-rev/2006.

\end{abstract}

\section{Introduction}

Meson spectroscopy has provided important clues to the understanding of the strong interactions. QCD is definitely well established as the basic theory, but we are still far from deriving the spectrum of mesons from first principles, including the determination of which quark configurations produce reasonably defined bound states and which do not. In this area, meson phenomenology can still provide valuable input for more fundamental considerations.

In this paper, we analyze the positive parity scalar mesons in the mass range 1-2 GeV. In the current picture, these states are supposed to  arise from P-wave, $q\bar q$ configurations, described long ago in~\cite{BorGa}. The well known, $J^{PC}=2^{++},~1^{++},~1^{+-}$ nonets, identified by $a_2(1320)$, $a_1(1260)$ and $b_1(1235)$ respectively~\cite{PDG} do fit in the $q\bar q$ picture quite well.
The case of  $0^{++}$ mesons is more complicated. The well-established, sub-GeV scalar mesons  $a_0(980)$ and $f_0(980)$ do not fit in the $(q\bar q)_{{\rm P-wave}}$ picture. In addition, lattice QCD predicts the appearence of a $0^{++}$ glueball at a mass $1500-1700$~ GeV~\cite{glueball}, followed by $2^{++}$ and $1^{+-}$ states at considerably higher mass. 

There are ten established scalar mesons in the 1-2 GeV range, namely: $a_0(1450)$, $K_0(1430)$, with $I=1,~1/2$ respectively, and three isosinglets: $f_0(1370)$, $f_0(1500)$ and $f_0(1710)$. It could be natural to identify these states with the $q\bar q$ nonet with $L=1$, $S=1$, $J=0$ ($L$, $S$, and $J$ denote orbital, spin and angular momentum of the $q\bar q$ system). This classification is however disfavored by two facts.  

i) Masses are too high. For $S=1$,  the spin-orbit coupling gives the ordering~\cite{BorGa}: $m(2^{++})\geq m(1^{++})\geq m(0^{++})$. Tensor forces may distort the regular spacing between multiplets, but are unlikely to push the scalar state on the top. In the case of charmonium states, for example, the scalar $\chi_0$ is  the lowest of $\chi$-states, with tensor forces pushing it  further down with respect to the spin-orbit coupling prediction.

ii)  Most important, $K_0(1430)$ is {\it lighter} than $a_0(1450)$ by some $62$~MeV, repeting the pattern of the $\kappa(800)$ with respect to $a_0(980)$ and at variance with the level ordering in the other positive parity nonets\footnote{In this paper we adopt the name convention of Particle Data Book; the number used to identify each particle deviates in some cases from the actual value of its mass, notably: $m[K_0(1430)]=(1412\pm6)$~MeV, $m[a_0(1450)]=(1474\pm19)$~MeV~\cite{PDG}.}.

Our starting point is that the reversed-octet pattern is natural for a tetraquark, $\left[qq\right]\left[{\bar q}{\bar q}\right]$ nonet with each pair antisymmetric in flavour. Such a structure has been advocated long ago  for the sub-GeV, scalar nonet~\cite{jaffeetal}, and was recently re-proposed~\cite{scalar1} in connection with new evidence for the $f_0(600)$ and $\kappa (800)$ resonances in D non leptonic decays~\cite{scalarlow}\cite{BES}, see also Ref.~\cite{Caprini} for recent theoretical developments. 

In the present paper,  we propose that the $0^{++}$ decimet is made by a tetraquark recurrence of the light $0^{++}$ nonet, with isosinglets mixed with the lightest glueball.  

We use a QCD inspired form of the mass matrix and of the decay amplitudes and neglect mixing with the lowest scalars.  We find that the picture  is consistent with the known information about the three $f_0$ resonances. We confirm the hypothesis~\cite{close} that the $f_0(1500)$ has a dominant glueball component. As a byproduct, we predict the mass of the still missing, $J^{PC}=1^{+-}$, singlet in the range $1470-1500$~MeV.

A crucial prediction of our model is that there should be an additional nonet, the genuine $q\bar q$, P-wave scalar, most naturally around or below the $a_1(1260)$.  In this region, an additional $I=1$ scalar state decaying in $K^{\pm} K^0_S$ was indicated time ago by  the OBELIX Collaboration~\cite{OBE98} and is not excluded by a more recent analysis of the $\eta\pi$ resonances in $p-\bar p$ annihilation~\cite{NWest}. There exists also some evidence for an additional isosinglet, e.g. the hypothetical $X(1110)$~\cite{other}. It seems fair to say that the possibility of a nonet around or below the $a_1(1260)$ cannot be excluded at the moment. 

The problem of scalar mesons has been addressed by several authors with very different proposals, see~\cite{PDGtornk} for a recent review. Here, we recall the investigations more closely related to our proposal.

In the framework of the $q\bar q$ picture, issues (i) and (ii) above have been addressed by Schecter and collaborators ~\cite{Schect}, who invoke a mixing of the higher and lower scalar nonets. 
The $f_0(1500)$ has been identified with a glueball mixed with the P-wave $q\bar q$ states $f_0(1370)$ and $f_0(1710)$  by Close and collaborators~\cite{close}. Spectrum and  mixing of scalar mesons with glueballs has been studied with QCD sum rules since the seminal investigation in Ref.~\cite{rubinstein}. In Ref.~\cite{kissling} a situation similar to that of Ref.~\cite{close} is found, namely a glueball mixed at 20\% with a $q\bar q$ meson\footnote{QCD sum rules have been applied to the correlation functions of pure glue operators and quark bilinears. It would be interesting to consider quark quadrilinears as well, to include tetraquark mesons in the picture.}. However, QCD sum rules give also a very light  glueball, in the range of the $f_0(600)$. Finally, the authors of Ref.~\cite{boglioni} argue that a single $q\bar q$ state could give rise to multiple resonances, e.g. to {\it both} $a_0(980)$ and $a_0(1450)$, due to dressing by hadronic interactions .

The variety of the proposed solutions points to the need of more accurate experiments to determine the actual number of scalar nonets as well as their decay properties, such as will be provided for $p-\bar p$ annihilation by the new antiproton facility at GSI.

The plan of the paper is as follows.

We give in Sect.~\ref{efflagr} the effective couplings used to analyse masses and decays of the mesons with negative parity nonets (identified by $\pi$ and $\rho$), positive parity nonets (identified by $a_0(980)$, $a_1(1260)$, $b_1(1235)$ and $a_2(1320)$) and the scalar decimet associated to $a_0(1450)$. We discuss the resulting mass spectra in Sect.~\ref{massspectr} and decay rates in Sect.~\ref{drsingl}, where we compare our results with Refs.~\cite{Schect} and~\cite{close}, and Sect.~\ref{droct}. 
We comment on the possible evidence for an additional scalar nonet in Sect.~\ref{compexpt}, conclusions are given in Sect.~\ref{conclusion}.

\section{ Effective lagrangians} 
\label{efflagr}
We describe nonet fields with the $3\times 3$ matrix, ${\bf \Pi}$: 
\be
{\bf \Pi}=\left(\begin{array}{ccc}\frac{X_1}{\sqrt{2}}+\frac{a_0^0}{\sqrt{2}} & a_0^+& K_0^+\\
a_0^- & \frac{X_1}{\sqrt{2}}-\frac{a_0^0}{\sqrt{2}}&K_0^0\\
K_0^- &{\bar K_0^0}&X_2\end{array}\right)
\label{mesonmatrix}
\ee
The states associated to $X_{1,2}$ correspond to the eigenstates of the quark mass matrix:
\ba
&&{\rm q\bar q:}\;\;\;\;\;\;\;\;\;\;\;\;X_1=|n\bar n>;\;\;X_2=|s\bar s> \nonumber  \\
&&{\rm \left[qq\right]\left[{\bar q}{\bar q}\right]:}\;\;\; X_1=|\left[ns\right]\left[{\bar n}{\bar s}\right]>;\;\;X_2=|\left[nn\right]\left[{\bar n}{\bar n}\right]>; 
\label{qqbar}
\ea
In the notation of Ref.~\cite{close}, $n$ represents $u$ and $d$ quarks, the isospin zero combination in $n\bar n$ and $nn$ is understood. 

We write the physical mass eigenstates according to:
\ba
&&f_{0,{\rm low}}=\cos\theta X_1+\sin\theta X_2\nonumber \\
&&f_{0,{\rm high}}=-\sin\theta X_1+\cos\theta X_2
\label{masseigen}
\ea
Interesting situations correspond to: $\theta \simeq 0$ ($q\bar q$, ideal mixing); $\theta \simeq 90^{\small 0}$ ($[qq][{\bar q}{\bar q}]$, ideal mixing); or to: $\theta \simeq -54^{\small 0}$ ($f_{0,{\rm low}}\simeq f_8$).

We parametrize the nonet mass matrix according to (${\bf \Delta}={\rm diag}(0,0,\delta)$):
\be
{\cal L}_{{\rm mass}}=m(a)+{\rm Tr}\left(
{\bf \Delta}\cdot {\bf \Pi}^2\right)+c[{\rm Tr}({\bf \Pi})]^2
\label{massa}
\ee
The second term represents the effect of the strange-to-light quark mass difference. The coefficient $c$ embodies the contribution of annihilation diagrams, with one quark pair transformed into any other one, via intermediate states made by pure glue.  If $\delta$ were dominant, the mass matrix would be diagonal in the $X_{1,2}$ basis, with $c$ dominant it would be diagonal in the octet-singlet basis. 

We expect annihilation to give a rather small contribution in the 1-2 GeV region, therefore we neglect SU$_3$ breaking in it. The mass formula Eq.~(\ref{massa}) is thus more restricted than required by a  first order expansion in $SU_3$ breaking. 

In the presence of a $0^{++}$ glueball we include additional terms mixing $X_{1,2}$ with the glueball as well as a diagonal glueball mass, $m_G=m(a)+\Delta m_G$. We neglect SU$_3$ breaking in the annihilation diagrams leading from quark matter to pure glue states, as before, so that the $X_{1,2}$ to glueball mixing is described by a single parameter, $w$.  In the $X_1, X_2, G$ basis, we are led to the $3\times 3$ matrix:
\be
{\bf M}=m(a)+\left(
\begin{array}{ccc} 2c&\sqrt{2}c&\sqrt{2}w\\
\sqrt{2}c&c+ 2\delta& w\\
\sqrt{2}w & w &\Delta m_G \end{array}\right)
\ee
We shall fix the combination $\Sigma=3c+\Delta m_G$ from the sum of isosinglet masses and take $\delta$ from the $K_0-a_0$ mass difference. Setting, in addition, $\Delta m_G=\Sigma(1-\alpha)$, we arrive at the 2-parameter formula:
\be
{\bf M}=m(a)+\left(
\begin{array}{ccc} \frac{2}{3}\alpha\Sigma &\frac{\sqrt{2}}{3}\alpha \Sigma&\sqrt{2}w\\
\frac{\sqrt{2}}{3}\alpha\Sigma& \frac{1}{3}\alpha\Sigma+ 2\delta& w\\
\sqrt{2}w & w &(1-\alpha)\Sigma \end{array}\right)
\label{mass3}
\ee
with: $\Sigma= \sum_i m(f^{(i)}_0)-2\delta-3 m(a)$, $\alpha$ and $w$ independent parameters.  The same formula with $\alpha=0$ was proposed in Ref.~\cite{cinese}. 

Next we consider decay rates. 

We restrict to decays into pairs of pseudoscalar, strictly octet, states. Amplitudes for the decay in isosinglet particles, i.e. $\eta$ or $\eta^\prime$, involve additional, undetermined parameters. In the exact $SU_3$ limit, the effective lagrangian contains three terms:

\be
{\cal L}_{decay}=A \{[ {\rm Tr} ({\bf \Pi}\cdot {\bf P}^2)- x\;{\rm Tr}({\bf \Pi}){\rm Tr}({\bf P}^2)]
+  \gamma\;{\bf G}\; {\rm Tr}({\bf P}^2)\}
\label{effectivel}
\ee 
where ${\bf P}$ is the pseudoscalar meson matrix, the analog of (\ref{mesonmatrix}), ${\bf G}$ the glueball field, $A$, $x$ and $\gamma$ unknown parameters, the latter describing the relative coupling of the glueball to two meson states. 

As a further simplification, we require that the quarks originally present in the decaying meson appear in the final mesons. It is an approximation related to the OZI rule~\cite{OZI} which forbids, e.g. $\phi(1020)\rightarrow \pi\pi$, and  is related to the smallness of annihilation processes. The rule restricts the value of $x$ by requiring that either $X_2$ (for $q\bar q$) or $X_1$ (for $[qq][\bar q\bar q]$) do not decay to $\pi \pi$. This gives: 
\ba
&&\rm{q\bar q:}\;\;\;\;\;\;\;\;\;\;\;\;x=0 \\
&&\rm{or}\nonumber \\
&&\rm{[qq][\bar q\bar q]:}\;\;\;\;x=-1/2\;\;\;\;\;\;\;\;\;\;   
\ea

For singlet decays, we have to substitute the combinations corresponding to the eigenstates of (\ref{mass3}), namely:
\be
f^{(i)}=c^{(i)}_1 X_1+c^{(i)}_2 X_2+c^{(i)}_3 {\bf G};\;\;   {\rm (i=1, 2, 3)}
\ee
If we call $g_{h}$ the coefficient in (\ref{effectivel}) of a given decay channel, the decay rate is:
\be
\Gamma_h=\frac{|\bf p|}{8\pi^3M^2}\;|F({\bf p}^2)|^2\;g_h^2\;
\label{scalwidth}
\ee
M is the mass of the decaying particle, $\bf p$ the decay momentum and $F({\bf p}^2)$ a form factor suppressing the emission of fast particles by the initial meson, due to its extended structure. In Ref.~\cite{close}, the parametrization is used:  
\be
|F({\bf p}^2)|^2=e^{-{\bf p}^2 / 8\beta^2}
\ee
with $\beta=400$~MeV. We shall consider also the extreme cases of no form factor ($\beta=+\infty$) or $\beta=240$~MeV, to exemplify a more extended structure than $q\bar q$.

\section{The mass spectrum} 
\label{massspectr}

For each nonet, the parameter $\delta$ is obtained from $K - a$ mass splitting and $c$ can be fixed by the value of the sum of isosinglet masses. Thus, we can predict the difference of the isosinglet masses, $\Delta m_f$, as well as the mixing angle. 

To test the accuracy of Eq.~(\ref{massa}), we have reported in Tables~\ref{subgev} and~\ref{abovegev} the predicted values of $\Delta m_f$ compared to the experimental ones, given in parentheses, for the well identified pseudoscalar, vector, tensor and axial vector nonets and for the two scalar multiplets. The $\delta$ parameters and the $X_1-X_2$ mixing angles are also reported. 
 
 Masses have been taken from~\cite{PDG} except for the lowest lying $0^{++}$ nonet, for which we  have taken the values quoted in Ref.~\cite{scalarlow}. The values quoted by the BES Collaboration~\cite{BES} are consistent, within the large experimental errors.  
 
 Our mass formula can be applied with squared masses. Quadratic mass formulae work somewhat better, as seen from the Tables, in particular for the pseudoscalar mesons. For simplicity of exposition, we shall restrict to linear mass formulae in what follows.
 
 \begin{table}
\begin{center}
\begin{tabular}{|c|c|c|c|c|c|c|}\hline
&$J^{PC}$  & $0^{-+} $ & $1^{--}$ & $0^{++}$          \\  \hline
&$\Delta m_f$~(MeV)    & 728(411) & 229(238) & 354(502) \\ \hline
{\bf lin. mass f.}&$\delta$~(MeV)   & 358& 117 & -188 \\ \hline 
&$\theta$   & -21$^\circ$& -2.1$^\circ$ & 79$^\circ$ \\ \hline \hline
&$\Delta m^2_f$~(GeV$^2$)     & 0.71(0.62) & 0.37(0.43) & 0.64(0.73) \\ \hline
{\bf quad. mass f.}&$\delta$~(GeV$^2$)    & 0.23 & 0.19 & -0.33 \\ \hline
&$\theta$  & -36$^\circ$&-4.4$^\circ$  &86$^\circ$  \\ \hline 
\end{tabular}            
\end{center}
\caption{Multiplets below 1 GeV. {\bf Above}, for linear mass formulae: i) predicted values of the isosinglets mass difference, $\Delta m_f$, compared to the  experimental values (in parentheses); ii) $SU_3$ breaking parameter, $\delta$; iii) isosinglet mixing angles. {\bf Below}: same for quadratic mass formulae. }
\label{subgev}
\end{table}

\begin{figure}[htb]
\begin{center}
\hspace{-3truecm}
\begin{minipage}[t]{40mm}
\epsfig{
height=4truecm, width=6truecm,
       figure=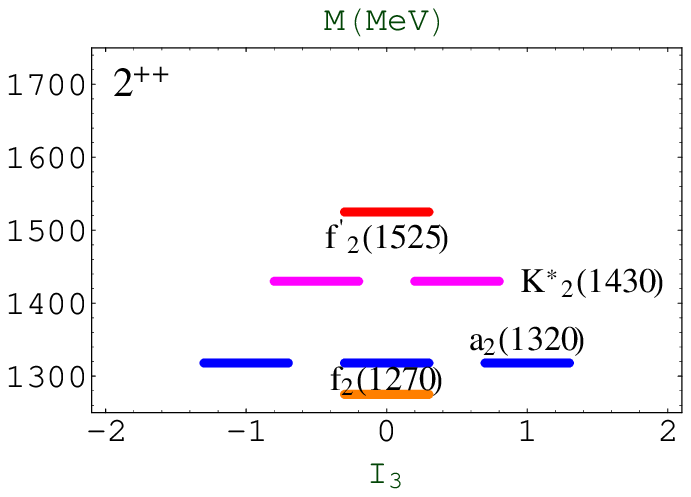}
\end{minipage}
\hspace{1.5truecm}
\begin{minipage}[t]{40mm}
\epsfig{
height=4truecm, width=6truecm,
        figure=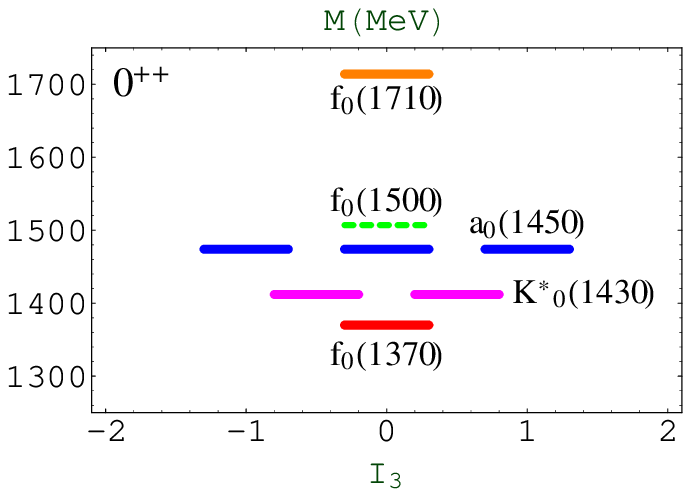}
\end{minipage}
\end{center}
\caption{\footnotesize  Mass spectra of the tensor mesons and of the higher scalar mesons. Same conventions of the previous figures. Note the inversion of the ordering of $I=1$ vs. $I=1/2$ states for scalar mesons, with respect to the well established $(q\bar q)_{S /P-wave}$ multiplets. In addition, there is an extra $0^{++}$, isoscalar, state, indicated by the green-dashed line.}

\label{tenscal}
\end{figure}

 The experimental spectra are illustrated in Fig.~\ref{belowGeV} to \ref{tenscal}. The difference of the pattern exhibited by the scalar multiplets sticks out conspicuously and is reflected in the negative sign of $\delta$ in Tables ~\ref{subgev} and~\ref{abovegev}.

A few comments are in order. 
\begin{itemize}
\item
With the well-known exception of the pseudoscalar mesons, the predicted values of $\Delta m_f$ are in general agreement with observation,  and mixing angles are all close to maximal, Eq.~(\ref{qqbar}), thus supporting the Ansatz made for the mass formula; 
\item 
Mixing angles reproduce well the affinities for non-strange ($\pi\pi$, $\rho\pi$, etc.) or strange ($K\bar K$, $K^*\bar K$, etc.) decay channels of the $q\bar q$ nonets and of the lowest scalar nonet, the latter under the $[qq][\bar q\bar q]$ hypothesis~\cite{scalar1};
\item
The contribution of gluon annihilation to the mass matrix is rather small as expected. As an example, we find for the the tensor mesons:
\be
\sqrt{2}c(2^{++})\simeq -28~\rm{MeV}.
\label{annih2}
\ee

\begin{figure}[htb]
\begin{minipage}[t]{40mm}
\epsfig{
height=4truecm, width=6truecm,
       figure=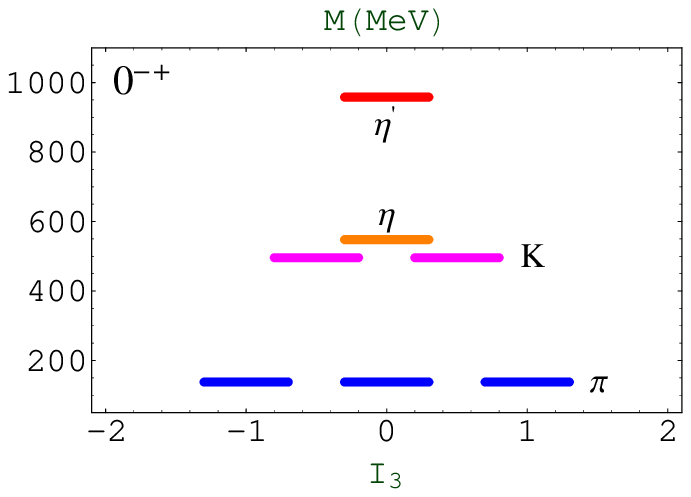}
\end{minipage}
\hspace{1.3truecm}
\begin{minipage}[t]{40mm}
\epsfig{
height=4truecm, width=6truecm,
        figure=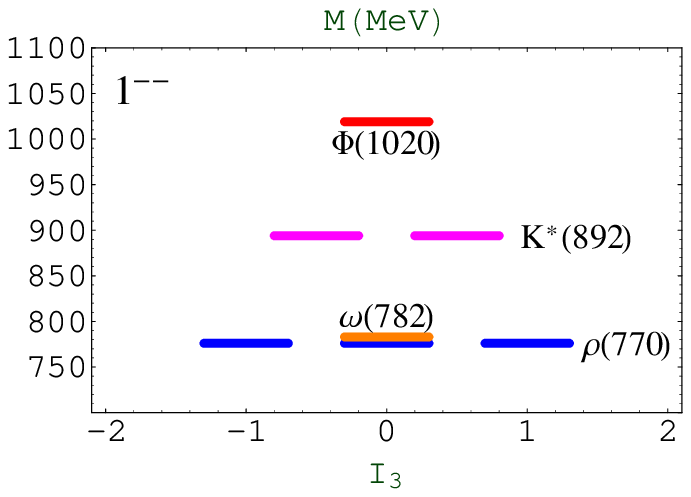}
\end{minipage}
\hspace{1.3truecm}
\begin{minipage}[t]{40mm}
\epsfig{
height=4truecm, width=6truecm,
        figure=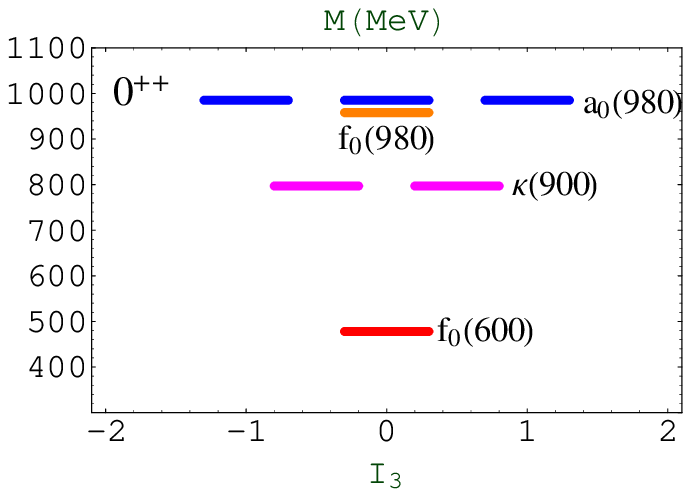}
\end{minipage}
\caption{\footnotesize  Mass spectra of the lowest lying meson multiplets, pseudoscalar, vector and the scalar mesons below 1 GeV. In blue the $I=1$ states, pink and orange $I=1/2$ and $I=0$ respectively. Masses and particle symbols are taken from Particle Data Group, Ref.~\cite{PDG}, except for $f_0(600)$ and $\kappa (800)$ where we use the values given by E791~\cite{scalarlow}.}
\label{belowGeV}
\end{figure}

\begin{table}
\begin{center}
\begin{tabular}{|c|c|c|c|c|c|c|c|}\hline
&$J^{PC}$  & $1^{++} $ & $1^{+-}$ & $2^{++}$ &$0^{++}$         \\  \hline
&$\Delta m_f$~(MeV)    & 144 (144 ) & 328 (--) &250  (250)  &  see text \\ \hline 
{\bf lin. mass f.}&$\delta$~(MeV)   & 67  & 145  & 112   &-62 \\ \hline
&$\theta$   & -24$^\circ$& 6.9$^\circ$ & 6.5$^\circ$ & see text  \\ \hline \hline
&$\Delta m^2_f$~(GeV$^2$)     & 0.39 (0.39) & 0.80 (--) & 0.67 (0.70) &  see text \\ \hline
{\bf quad. mass f.}&$\delta$~(GeV$^2$)    & 0.19  & 0.36  &0.31  &-0.18 \\ \hline
&$\theta$  &  -20$^\circ$& 6.7$^\circ$  & 5.6$^\circ$ &  see text  \\ \hline  
\end{tabular}            
\end{center}
\caption{Same as in Table~\ref{subgev} for the multiplets in 1-2 GeV. For the $1^{+-}$ the experimental $\Delta m_f$ is not available since $h^\prime_0$ has not yet been observed.}
\label{abovegev}
\end{table}

 \item
 The two axial vector nonets differ in charge conjugation but K-states are mixed by $SU_3$ breaking. The mass matrix contains seven parameters which can be determined from the known masses, yielding a prediction for the yet unknown $h^\prime_{1}$, with a slight difference between linear and quadratic mass formula: 
\be
m(h^\prime_{1})= 1474~{\rm MeV}\;{\rm (quadratic)};\;\; m(h^\prime_{1})=1498~{\rm MeV}\;{\rm (linear)}
\ee
To fit the masses, we need a rather small non diagonal element of the Hamiltonian between Kaon states: 
\be
<K^{++}|H|K^{+-}> = 53~\rm{MeV}
\label{mix1}
\ee

 \end{itemize}

The results in (\ref{annih2}) and (\ref{mix1}) indicate rather small configuration mixing and annihilation contributions in the meson states above one GeV.

\begin{figure}[htb]
\begin{minipage}[t]{40mm}
\epsfig{
height=4truecm, width=6truecm,
       figure=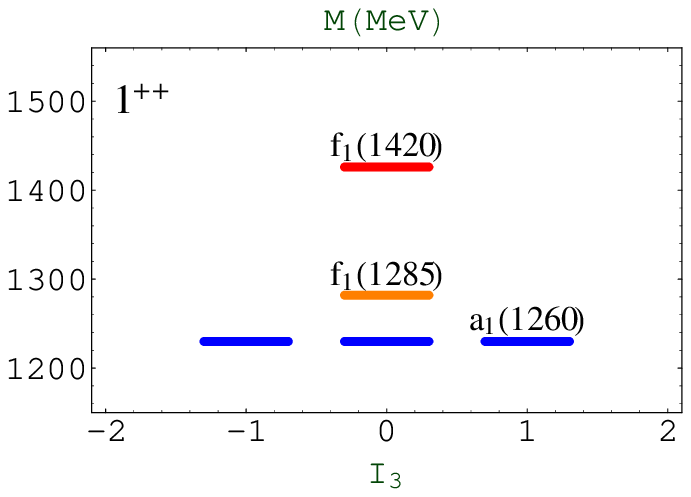}
\end{minipage}
\hspace{1.3truecm}
\begin{minipage}[t]{40mm}
\epsfig{
height=4truecm, width=6truecm,
        figure=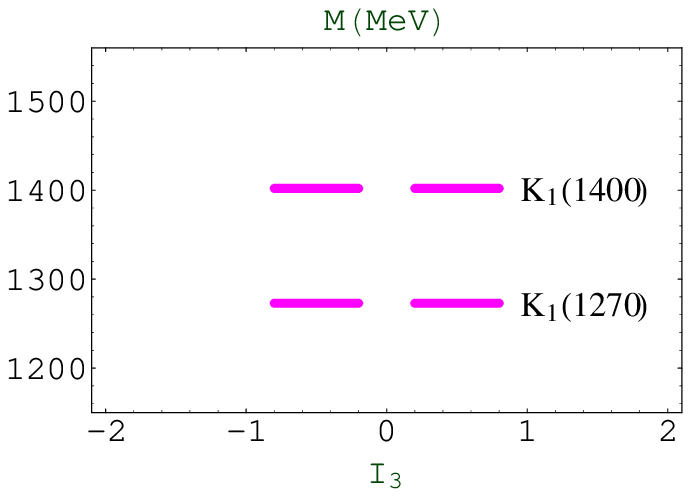}
\end{minipage}
\hspace{1.3truecm}
\begin{minipage}[t]{40mm}
\epsfig{
height=4truecm, width=6truecm,
        figure=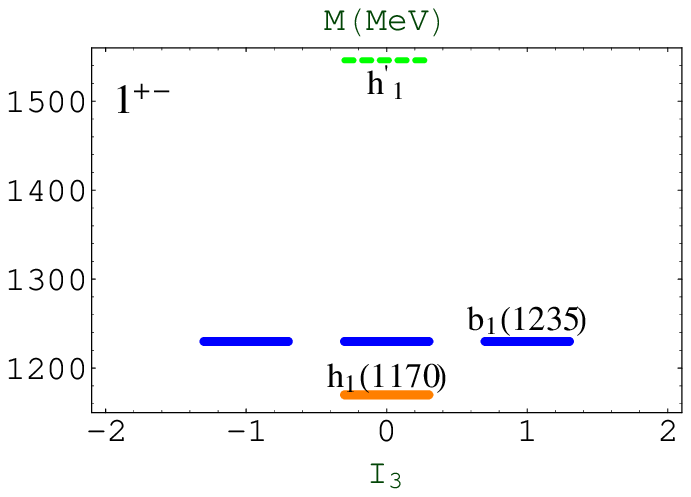}
\end{minipage}
\caption{\footnotesize  Mass spectra of the axial mesons multiplets, same conventions as in Fig.~\ref{belowGeV}. Strange states are mixed by SU$_3$ breaking and are reported separately. The green-dashed line indicates the predicted position of $h^{\prime}_1$, yet to be observed state.}
\label{aximes}
\end{figure}

For the $0^{++}$ decimet, we solve numerically  the eigenvalue equation corresponding to Eq.~(\ref{mass3}).  We use masses and errors given in~\cite{PDG} to compute the $\chi^2$  of each determination ($\chi^2=\sum_i(m_{{\rm th}}-m_{{\rm exp}})^2/\Delta m_{{\rm exp}}^2$) except for $f_0(1370)$ whose mass is rather poorly known. In this case we take, for definiteness: $m= 1370\pm 60$~MeV. 

\begin{figure}[htb]
\begin{center}
\epsfig{
height=5truecm, width=8truecm,
        figure=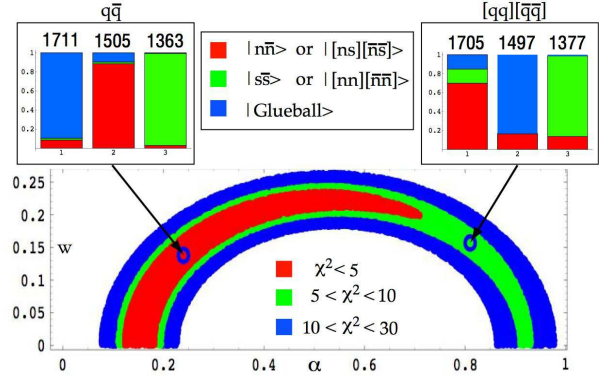}
\caption{\footnotesize Distribution in parameter space of the $\chi^2$ computed from the eigenvalues of the mass matrix~(\ref{mass3}). The arrow in the right panel points to the region preferred by particle decays in the $[qq][\bar q\bar q]$ hypothesis. The bars in the panel give the probabilities of quark/glueball composition of the eigenvectors. The left panel corresponds to the $q\bar q$ hypothesis. }
\label{rainbow}
\end{center}
\end{figure}

We show in Fig.~\ref{rainbow} the distribution of $\chi^2$ in parameter space\footnote{$w$ can be made to be positive by an appropriate definition of the sign of the G field}. The red region gives a better $\chi^2$, but acceptable masses are found in all the green region. In the left corner ($w\simeq 0$, $\alpha$ small) the glueball corresponds to the highest mass and it moves to the intermediate mass at the right corner ($w\simeq 0$, $\alpha$ large). Of the other two states, the highest one is mostly $X_1$ and the lowest $X_2$. For $q\bar q$ states this corresponds to the $s\bar s$ state being lighter than the $n\bar n$ and it is, of course, a consequence of the unnatural sign of the effective strange-to-light quark mass difference, implied by the $K_0-a_0$ splitting. For tetraquark states, on the contrary, this is the natural level ordering, Eq.~(\ref{qqbar}).

\section{Decay rates of singlet mesons and best solution} 
\label{drsingl}

For singlet meson decays, we have explored the parameter space by optimizing the $\chi^2$ with respect to the ratios of branching fractions reported in the first two columns of Table~\ref{t:brfrac}. 

In the $[qq][\bar q\bar q]$ hypothesis, a reasonable picture is obtained for the point indicated in Fig.~\ref{rainbow} right panel, with: $\alpha=0.81$, $w/\Sigma=0.16$ and $\gamma=-0.26$.

For reference, we give explicitly the mass matrix (\ref{masstetraq}), in the basis $X_1$, $X_2$, $G$:
\ba
&& M=\left( \begin{array}{ccc}1626 &108 & 64 \\ 108 &  1426 &45 \\64 & 45 &1528 \end{array}\right)
\label{masstetraq}
\ea
and the corresponding eigenvectors and eigenvalues:
\ba
&&\left(\begin{array}{c} f^{(1)}(1706)\\f^{(2)}(1497)\\f^{(3)}(1377)\end{array}\right)= \left( \begin{array}{ccc}-0.834 & -0.385 & -0.395  \\ -0.411 &  -0.0447 &0.911 \\0.368 & -0.922&0.121 \end{array}\right) \left(  \begin{array}{c} X_1 \\ X_2 \\ G \end{array} \right)
\label{4qmix}
\ea
The value $m_G= 1528$ is consistent with lattice determinations~\cite{glueball}. 

The numerical values of the branching fraction ratios  computed in correspondence to Eq.~({4qmix}) are given in Table~\ref{t:brfrac}. Values generally agree with the trend of the experimental figures. Larger form factor damping is preferred. Values in the third column reproduce well the affinity of $f_0(1370)$ to non strange channels (indicated independently by the large branching ratio in four pions: $B_{4\pi}[f_0(1370)] \simeq 80\%$~\cite{PDG}). 

The same procedure with the $q\bar q$ hypothesis gives as best solution the point indicated by the left panel in Fig.~\ref{rainbow}, corresponding to $\alpha=0.23$, $w/\Sigma=0.14$. The mass matrix and corresponding eigenvalues and eigenvectors for this case are:
\ba
&& M=\left( \begin{array}{ccc}1519& 32& 56 \\ 32 & 1373 & 39 \\ 56 & 39 & 1688 \end{array}\right)
\label{mass2q}
\ea

\ba
&&\left(\begin{array}{c} f^{(1)}(1711)\\f^{(2)}(1505)\\f^{(3)}(1363)\end{array}\right)= \left( \begin{array}{ccc} 0.294 &  0.136 &  0.946  \\-0.941&  -0.135 &  0.312 \\0.170 & -0.981 & 0.089 \end{array}\right) \left(  \begin{array}{c} X_1 \\ X_2 \\ G \end{array} \right)
\ea   
   
Branching fraction ratios for the $q\bar q$ assumption are also reported in Table~\ref{t:brfrac}. For the first two columns we find reasonable agreement with data. However, within our model of the mass matrix, the lowest state is essentially $s\bar s$, see Fig.~\ref{rainbow} and the $f_0(1370)$ is predicted to have a strong affinity for strange decay channels, in contrast with data.

\begin{table}[h]
\begin{center}
\begin{tabular}{|c|c|c|c|c|}\hline
& & ${\cal B}_{\pi\pi}/{\cal B}_{KK} [f_0(1710)] $ & ${\cal B}_{KK}/{\cal B}_{\pi\pi}[f_0(1500)] $ &${\cal B}_{KK}/{\cal B}_{\pi\pi}[f_0(1370)] $ \\  \hline
& no form factor & 0.5 & 0.18 &0.14 \\ \hline
$[qq][\bar q\bar q]$& $\beta=$400~MeV~\cite{close} & 0.42 & 0.23 &0.17\\ \hline
& $\beta=$240~MeV & 0.31 & 0.32& 0.23 \\ \hline
$q\bar q$& no form factor & 0.24 & 0.42 & 22\\ \hline
Expt. & & $<0.11^{a}$, $0.24^{b} \pm 0.024 \pm 0.036$ & $0.246\pm 0.026$ & 0.1$^c - 0.9^d$\\ \hline
\end{tabular}            
\end{center}
\caption{Ratios of branching fractions for $\pi\pi$ and $K\bar K$ decays of the $0^{++}$ isosinglets. Above: $[qq][\bar q\bar q]$ hypothesis with three different assumptions on the form factor. Middle: $q\bar q$ hypothesis (no form factor gives the best agreement). Below: experimental data from {\it Particle Listing 2005}~\cite{PDG}: a~\cite{ablikim04}, b~\cite{barberis99}, c~\cite{ablikim05}, d~\cite{bargiotti03}. }
\label{t:brfrac}
\end{table}

Before closing this Section, we compare our results with the previous proposals in~\cite{Schect} and~\cite{close}. 

In~\cite{Schect}, the anomalous properties of the $0^{++}$ multiplet above 1 GeV are ascribed to a mixing of the P-wave $q\bar q$ states with the the sub-GeV scalar mesons, assumed to be of a different kind (i.e. hadronic molecules or tetraquarks). A suitable mixing of the two multiplets  could push the $q\bar q$ scalars above what expected in the spin-orbit coupling scheme and in addition make the $I=1$ state heavier than the $I=1/2$ one.  

The mixing needs to be very fine-tuned and, in addition, it turns out to be quite large. Quantitatively, for linear mass-formulae, a non diagonal matrix element of the Hamiltonian of about $250$~MeV is required in Ref.~\cite{Schect} to push the $a_0(1450)$ above the $K_0(1430)$. 
The result is rather unnatural given the very different  configurations  of the states involved ($q\bar q$ vs. $qq\bar q\bar q$) and that it involves quark pair annihilation. In the more favourable case of the strange mesons belonging to the $1^{++}$ and $1^{+-}$ octets (both $q\bar q$ configuration, no annihilation required) we have found a quite smaller result, Eq.~(\ref{mix1}). On our side, the non diagonal matrix elements in Eqs.~(\ref{masstetraq}, \ref{mass2q}), are similar to (\ref{mix1}).

Comparison with ~\cite{close} requires some qualification.
Without mixing to the sub-GeV multiplet, the inverted pattern of the $I=1$ and $I=1/2$ masses could still be explained, in the $q-\bar q$ model, as due to an overcompensation of the strange to non strange quark mass difference by spin effects (which also depend upon quark masses). Within our Ansatz for the mass matrix, however, the wrong sign of the SU$_3$ breaking in the octet propagates to the singlets and makes it so  that the scalar made by strange quarks is {\it lighter} than the one made by non strange quarks. For this reason,  the $q\bar q$+glueball assumption fails in reproducing the affinity of $f_0(1370)$ for non strange channels, Table~\ref{t:brfrac}. 
The tetraquark+glueball model fares better in that the association of the heavier meson with strange quarks is preserved in the reversed octet. 

With a more general mass matrix, the correlation between octet and singlets would be lost and the observed affinities to non strange or strange channels could be reproduced. The difficulty would remain, however, to explain the opposite sign of SU$_3$ breaking in the $0^{++}$ octet with respect to the other P-wave octets.

\section{The decay rates: $a_0(1450)\rightarrow K\bar K$ and $K_0(1430)\rightarrow K\pi$}
\label{droct}

In the exact SU$_3$ limit, these partial decays are given in terms of one single amplitude. Comparing with data, we can test wether the two particles belong indeed to the same SU$_3$ octet. 
Ref.~\cite{PDG} quotes:
\ba
&&\Gamma[K_0(1430)]= 294\pm 23~{\rm{MeV}}\nonumber \\
&&B[K_0(1430)\rightarrow K\pi]=0.93\pm 0.10
\ea
whereby:
\be
\Gamma[K_0(1430)\rightarrow K\pi]=253\pm 51~{\rm{MeV}}
\ee
Using (\ref{scalwidth}), in the SU$_3$ limit we predict:
\be
\Gamma (a_0^+\rightarrow K\bar K)_{SU_3}=\frac{2}{3}\frac{p_{a_0}}{p_{K_0}}\frac{M_{K_0}^2}{M_{a_0}^2}\Gamma (K_0^+\rightarrow K\pi)=150\pm 28~{\rm{MeV}}
\ee

As for data on $a_0(1450)$ decays, Ref.~\cite{PDG} quotes three observed decay channels: $\eta\pi$, $\eta^\prime\pi$ and $\omega\pi\pi$, and gives total width and ratios of the partial widths to the $\eta\pi$ one:
\ba
&&\Gamma(a_0(1450))=265\pm 13~{\rm{MeV}};\nonumber \\
&&\frac{\Gamma(\eta^\prime\pi)}{\Gamma(\eta\pi)}=0.35\pm 0.16;\;\;\frac{\Gamma(K\bar K)}{\Gamma(\eta\pi)}=0.88\pm0.23;\;\;\ \frac{\Gamma(\omega\pi\pi)}{\Gamma(\eta\pi)}={\rm{preferred\;\;value\;\;not \;\;quoted}}
\ea

Neglecting the $\omega\pi\pi$ contribution, we find:
\be
\Gamma (a_0^+\rightarrow K\bar K)=105\pm46~{\rm{MeV}}
\ee

In~\cite{Schect} the discrepancy of the central values was taken to support unequal mixing angles of $a_0(1450)$ and $K_0(1430)$ with their lighter partners. However, this would be an SU$_3$ breaking effect of the same order of similar effects neglected in decay rates, so the argument may not be so compelling. In fact, given the large experimental errors and the neglect of SU$_3$ breaking, the assignment to the same multiplet can be consistent.

A reduction of the experimental errors and, most important, a determination of the partial width of $a_0(1450)\rightarrow \omega\pi\pi$ would greatly help to clarify the situation.

\section{Another scalar nonet?}
\label{compexpt}

The crucial feature of our model is that it requires an additional nonet, the genuine $q\bar q$, P-wave scalar, around or below the $a_1(1260)$.  
Experimental searches for SU$_3$ scalar nonets have been focussed on isovector resonances decaying into $K^\pm K^0_S$ or $\eta\pi$.  

The very accurate analysis of $p-\bar p$ annihilation at rest done by the OBELIX Collaboration~\cite{OBE98} showed a clear peak in the mass distribution of  $K^\pm K^0_S$ around 1.3 GeV, where the $a_2(1320)$ is expected. However, a comparison of annihilation events produced in targets with different density showed that the peak shifts from 1.32 GeV at the lowest density target (where most of the reaction proceeds from P-wave), to 1.29 GeV in the liquid Hydrogen target (where annihilation from S-wave dominates). The best fit for the 1.29 GeV resonance is obtained for $J^{PC}=0^{++}$, indicating an additional scalar isovector $a_0(1290)$, produced dominantly by S-wave annihilation.

The most recent, high-statistics, analysis presented in~\cite{NWest} used $p-\bar p$ annihilation in flight, where it is not possible to separate the different partial waves, and the $\eta\pi$ channel. Also in this case, a peak around 1.3 GeV was seen. The best fit was indeed obtained for $J^{PC}=0^{++}$,  with mass consistent with $a_0(1290)$ but a larger width than indicated by OBELIX. A lower quality fit is obtained for the $a_2(1320)$, which was preferred ``because of the weight of numerous final states, including $\eta\pi$ , in which $a_2(1320)$ has been observed, and not $a_0(1290)$". In view of the OBELIX data, the most likely situation would be in fact that the two resonances coexist in the final states. This hypothesis was also tried in~\cite{NWest} but with inconclusive results.

It seems to us fair to say that the possibility of a nonet around or below the $a_1(1260)$ cannot be excluded at the moment and that the issue of the $a_0(1290)$, a good candidate for the $q\bar q$ P-wave scalar, deserves a closer scrutiny.

\section{Conclusions}
\label{conclusion}

Scalar mesons present many puzzling aspects, which have given rise to several different speculations. 

We have presented a novel proposal in which the presently observed scalars, both below 1 GeV and in the  1-2 GeV range, are tetraquark states. This explain naturally the mass ordering of $I=1/2$ and $I=1$ states exhibited by both multiplets and is consistent with known facts about their decays. The $f_0(1500)$ is naturally included in the picture as a light glueball, mixed with the tetraquark singlet states $f_0(1370)$ and $f_0(1710)$.

A direct consequence of our proposal is that there should exist another scalar nonet, probably around 1.2 GeV, the genuine $q\bar q$ P-wave state. Evidence of such a complex may have been seen by OBELIX in $p\bar p$ annihilation at rest and may be indicated, or anyway not excluded, by a recent  high statistics experiment on $p\bar p$ annihilation at higher energy.

A thorough exploration of the scalar spectrum in the 1-2 GeV region and a more precise determination of decay modes and branching ratios is clearly needed for the resolution of the nature of scalar mesons, an issue of the utmost importance for understanding the way QCD works for bound states.

\vskip0.5cm
{\bf {\sl Acknowledgments}}. 
We thank F. Close for interesting exchange on the properties of the $f_0$(1500) and U.~Gastaldi, A.~Masoni and C.~Voena for useful information on the experimental data.

\end{document}